\documentclass[aps,prl,twocolumn,epsfig,groupedaddress]{revtex4}
\usepackage{epsfig}
\usepackage{units}

\begin{document}

\title{A local Monte Carlo implementation of the non-abelian Landau-Pomerantschuk-Migdal effect}

\author{Korinna Zapp$^{(1,2)}$,
Johanna Stachel$^{(1)}$
Urs Achim Wiedemann$^{(3)}$}

\address{$^{(1)}$Physikalisches Institut, Universit\"at Heidelberg, Philosophenweg 12, D-69120 Heidelberg, Germany\\
$^{(2)}$EMMI, GSI Helmholtz-Institut f\"ur Ionenforschung, Planckstr. 1, D-64291 Darmstadt\\
$^{(3)}$Physics Department, Theory Unit, CERN, 
CH-1211 Gen\`eve 23, Switzerland}

\date{\today}

\begin{abstract}
The non-abelian Landau-Pomeranschuk-Migdal (LPM) effect arises from the quantum 
interference between spatially separated, inelastic radiation processes in matter. A consistent probabilistic implementation of this LPM effect is a prerequisite for extending the use of Monte 
Carlo (MC) event generators to the simulation of jet-like multi-particle final states in nuclear 
collisions. Here, we propose a local MC algorithm, which is based solely on relating the LPM 
effect to the probabilistic concept of formation time for virtual quanta. We demonstrate that
this implementation of formation time physics alone accounts probabilistically for all
analytically known features of the non-abelian LPM-effect, including the characteristic $L^2$-dependence of average parton energy loss and
the characteristic $1/\sqrt{\omega}$-modification of the gluon energy distribution.
Additional kinematic constraints are found to modify these $L^2$- and $\omega$-dependencies characteristically in accordance with analytical estimates. 
\end{abstract}

\maketitle



In heavy ion collisions, highly energetic partons propagate through 
the produced dense QCD matter before reaching the QCD vacuum. 
The fragmentation of these partons is modified significantly by the medium. This is
seen in the strong suppression of essentially all hadronic spectra at high transverse
momentum, discovered at RHIC~\cite{RHIC}, and is consistent with calculations of 
medium-induced parton energy loss~\cite{Baier:1996sk,Zakharov:1997uu,Wiedemann:2000za,Gyulassy:2000er,Arnold:2002ja,Wang:2001ifa}. Heavy ion collisions at the 
LHC will open qualitatively novel opportunities for studying this jet quenching 
effect~\cite{LHC}, since jet-like multi-hadron final states will become experimentally 
accessible in a wide kinematical regime, which is well-separated from the
soft high-multiplicity features of nuclear collisions. To exploit these novel
opportunities fully, a reliable theoretical framework for the description of multi-particle
final states and their medium-modification is needed. In the absence of a medium,
the standard tool is the MC simulation of the final state parton shower  
underlying jet fragmentation. Its dominant medium-modification is expected to be determined
by the non-abelian LPM effect. The question arises, how this medium-induced quantum interference
effect can be implemented in a probabilistic MC simulation. Here, we propose a MC
algorithm to this end. 

In the vacuum, a virtual parton $p=q,g$ of energy $E$
degrades its virtuality mainly by radiating a gluon of energy $\omega$ and transverse
momentum $k_{\perp}$. The gluon energy distribution is 
$ \frac{dI^{(vac)}}{d\omega\, dk_{\perp}^2} \sim \frac{\alpha_s}{E}\, P_{g/p}(z)\, \frac{1}{k_{\perp}^2}$,
where $P_{g/p}(z) \sim \frac{1}{z} + O(z^0)$, $z= \omega/E$ denotes the standard $p \to p+ g$
splitting function in the vacuum. The consequences of embedding the parton splitting process
$p \to p + g$ into a medium have been studied in detail~\cite{Baier:1996sk,Zakharov:1997uu,Wiedemann:2000za,Gyulassy:2000er,Arnold:2002ja,Wang:2001ifa}. For relativistic
energy, the leading medium-modification can be written in terms of an 
additive term, $\omega \frac{dI^{(med)}}{d\omega } \propto \alpha_s 
\sqrt{\frac{\omega_c}{\omega}}$, which shows the characteristic $1/\sqrt{\omega}$-
dependence of the non-abelian Landau-Pomerantschuk-Migdal (LPM) effect. This
contribution persists up to an energy scale $\omega < \omega_c = \frac{1}{2} \hat{q}\, L^2$,
defined in terms of the in-medium path length $L$ and the quenching parameter
$\hat{q}$~\cite{Baier:1996sk,Zakharov:1997uu,Wiedemann:2000za}. 

There have been several MC approaches to the problem of radiative
parton energy loss. These implement the LPM effect for instance by some form of
parametrized suppression of the projectile energy~\cite{Lokhtin:2005px} or a
parametrized modification of the parton splitting~\cite{Zapp:2008gi,Armesto:2007dt}.
Destructive interference in inelastic processes has also been modeled by assigning to
newly produced partons a mean free path, during which no further interactions can 
occur~\cite{Xu:2007jv}. Moreover, medium-induced parton energy loss has been
recognized to change the virtuality evolution of the parton shower 
(see e.g.~\cite{Wang:2001ifa,Majumder:2009zu}) and there is a MC model implementing
a medium-modified virtuality evolution~\cite{Renk:2008pp}. 
The MC algorithm proposed here differs from these approaches in that the LPM
effect does not emerge as a consequence of a parametrized modification of the parton 
cascade. Rather, it is the dynamical result of a local, energy-momentum conserving
evolution, which is consistent with analytically known limiting cases of
~\cite{Baier:1996sk,Zakharov:1997uu,Wiedemann:2000za,Gyulassy:2000er,Arnold:2002ja,Wang:2001ifa}.

To understand how LPM quantum interference can be built up in a local probabilistic
dynamics involving multiple scattering, we recall first analytical results in known
limiting cases~\cite{Wiedemann:2000za}. We start with the case of 
a gluon radiation process involving two transverse momentum transfers $q_{\perp,1}$ and $q_{\perp,2}$ within path length $L$. The resulting distribution of radiated gluons is known
analytically for $E \gg \omega \gg \vert k_\perp\vert$. It interpolates between two 
limiting cases. In the totally incoherent production limit, the gluon is produced in the first 
scattering and its transverse momentum is shifted probabilistically by $q_{\perp,2}$ 
in the second scattering. In the opposite limit, the two scattering centers are not resolved 
and the radiated gluon spectrum corresponds to that of a single (effective) scattering center,
in which the momentum $(q_{\perp,1} + q_{\perp,2})$  was exchanged with the 
target~\cite{Wiedemann:2000za}. The transition between these limits occurs at a
scale $L \sim 1/Q_1 \equiv 2\omega/ \left( k_{\perp,1}\right)^2$, where
$k_{\perp,1}$ is the transverse momentum of the radiated gluon after
the first  momentum transfer. This scale enters the analytical result via the
interference factor  $2 \left(1 - \cos Q_1\, L \right)/\left( Q_1\, L\right)^2$,
which vanishes in the incoherent limit $L \gg 1/Q_1$ and which
approximates unity in the totally coherent limit $L \ll 1/Q_1$~\cite{Wiedemann:2000za}. 
The term $1/Q_1$ can be interpreted as the 
gluon formation time after the first scattering. If the second scattering occurs within the
gluon formation time, then it acts coherently, otherwise incoherently. 
By inspecting analytical results for gluon production processes involving more than two 
momentum transfers~\cite{Wiedemann:2000za}, one finds that this role of the inverse 
transverse gluon energy as formation time is generic~\cite{tobedone}: 
the next $(n+1)$-th scattering center contributes totally coherently  (incoherently) to a gluon 
production process, if the gluon formation time  
\begin{equation}
   t_f = \frac{2\omega}{k_{\perp,n}^2}
   \label{form}
\end{equation}
after the $n$-th scattering is much larger (much shorter) than the distance $d$ to the 
$(n+1)$-th scattering center. 

These limiting cases of totally coherent and incoherent gluon production are implemented
easily in a probabilistic MC algorithm. Consider a target scattering center $Q_T$ 
participating in an inelastic process. In the incoherent case, the gluon will be produced
on this scattering center with inelastic cross section $\sigma^{qQ_T \to qQ_T g}$.
Subsequent scatterings can change with elastic cross section $\sigma^{gQ_T \to gQ_T}$ the 
transverse momentum of the fully formed gluon to $k_T \to k_T + \sum_i q_{\perp,i}$,
but they are not necessary to complete the gluon radiation process.
In the opposite limit of totally coherent gluon production, all $n$ scattering 
centers within the path-length $L$ transfer their individual momenta $q_{\perp, i}$
coherently to the projectile, as if the gluon were radiated in a 
single inelastic scattering process with transverse momentum transfer
\begin{equation}
	q_{\perp, tot} = \sum_{i=1}^N q_{\perp, i}\, .
\end{equation}
Although this limit is dominated by quantum interference, it can be realized exactly in a 
MC algorithm by selecting probabilistically 
those target scattering centers, which transfer a non-vanishing transverse momentum 
$q_{\perp, i}$ from the medium to the projectile. The distribution of the $q_{\perp, i}$ is
determined by the elastic cross section. 
The gluon is then generated according to the cross section  $\sigma^{qQ_T \to qQ_T g}$ 
but with the constraint that the momentum transfer from the target
is $q_{\perp, tot}$. This procedure matches exactly the result of analytical 
calculations~\cite{Wiedemann:2000za}. 

In between the totally coherent and incoherent limits, 
we propose that {\it the LPM effect can be implemented in a probabilistic MC algorithm 
by requiring that the momentum transfer from different scattering centers to the partonic 
projectile acts totally coherently for gluon production, if it occurs within the formation time 
$t_f$, and that it acts incoherently, if it occurs after $t_f$.} This is implemented
as follows: 
Once a scattering center is selected as the source of an inelastic gluon production process,
the MC determines probabilistically the distance $d$ to the next scattering
center with which the projectile (quark and gluon) interacts. If $d > t_f$, then the gluon is 
regarded as formed, and it is treated in subsequent steps of the dynamical evolution as an 
independent projectile. If $d < t_f$, then the momentum transfer $q'_\perp$ of the subsequent 
scattering center is added coherently to the one in the gluon production process.
This changes the gluon transverse momentum,  and thus the gluon formation time to
$t'_f \simeq \frac{2\omega}{(k_\perp + q'_\perp)^2}$. Then, if the next interaction lies 
outside the length $t'_f$, the gluon is regarded as formed.  Otherwise, one repeats the above procedure,  until the gluon is formed. Another inelastic interaction is only generated once the 
gluon is fully formed. This defines a local, probabilistic time evolution, which 
reproduces by construction the analytically known totally coherent and incoherent limiting 
cases. 

We now demonstrate in a numerical model study that such a MC algorithm reproduces 
the main characteristic features of the LPM effect. To be specific, we model the 
target as an ensemble of colored scattering
centers $Q_T$ with density $n$. The scattering centers
present elastic and inelastic cross sections 
$\sigma^{qQ_T\to qQ_T}$  and $\sigma^{qQ_T \to qQ_T g}$ to a projectile parton.
Here, we use the ansatz $\frac{d}{d\omega}
\sigma^{qQ_T \to qQ_T g} = g^2\, C_F\, \frac{1}{\omega}\, \sigma^{qQ_T\to qQ_T}$,
which was used in the calculations of ~\cite{Baier:1996sk,Zakharov:1997uu,Wiedemann:2000za,Gyulassy:2000er,Arnold:2002ja,Wang:2001ifa}. 
The results shown below were obtained for total elastic and total inelastic cross sections 
fixed such that $n\, \sigma^{qQ_T \to qQ_T g} = 1\, {\rm fm}^{-1}$ and 
$n\, \sigma^{gQ_T \to gQ_T } = 1 \,{\rm fm}^{-1}$. This correspond to 
$\hat{q} \simeq 1\, {\rm GeV}^2/{\rm fm}$, where $\hat{q}$ characterizes 
the strength of elastic scatterings 
\begin{equation}
	\hat{q} = n\, \int dq_{\perp}^2\, q_{\perp}^2\, 
		\frac{d \sigma^{qQ_T \to qQ_T } }{dq_\perp^2}\, .
\end{equation}
We checked that consistent with analytical results~\cite{Baier:1996sk}, 
the $\hat{q}$- and $L$-dependence of the MC simulated 
spectrum $\omega \frac{dI}{d\omega}$ can be accommodated in a 
function of $\omega/\omega_c$ (data not shown). As a consequence,
our numerical results do not rely on a particular parameter choice for $\hat{q}$ and
we plot them in suitably rescaled dimensionless quantities.
%
\begin{figure}[h]
\begin{center}
\includegraphics{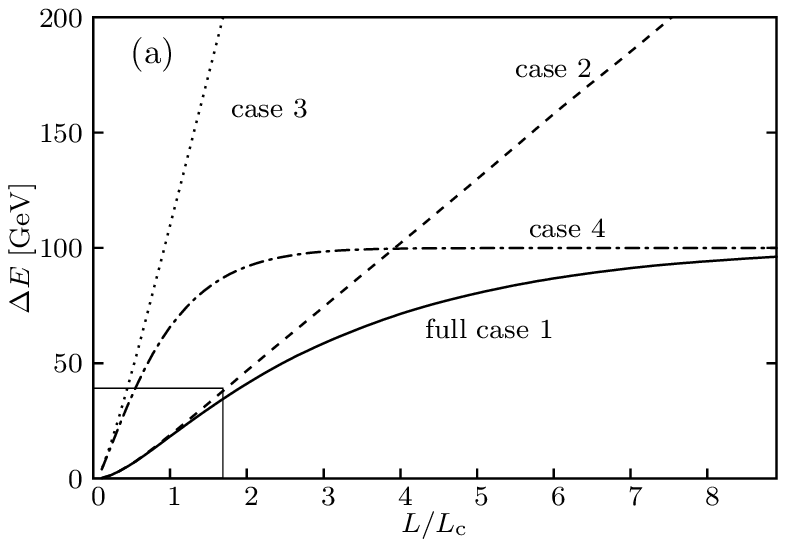}
\includegraphics{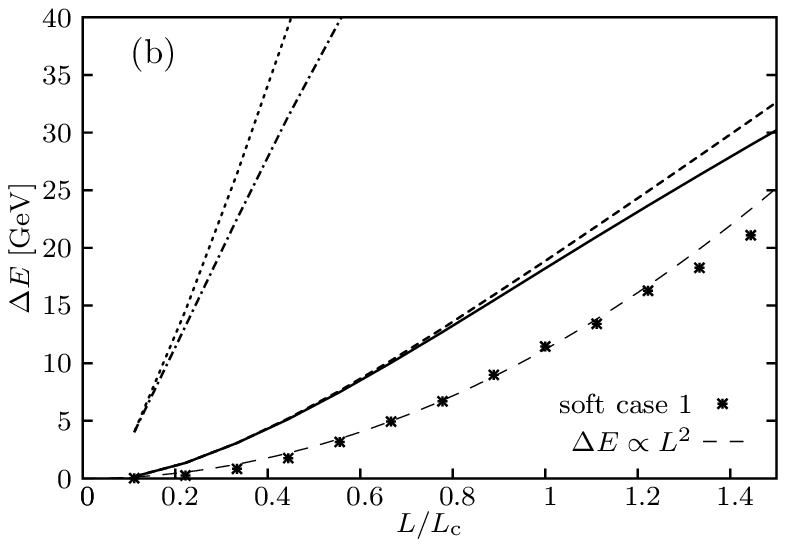}
\end{center}
\vspace{-0.5cm}
\caption{The medium-induced energy loss $\Delta E$ of a quark of initial
energy $E_q = 100\, {\rm GeV}$ as a function of the in-medium path length $L$.
Results for the full MC algorithm with LPM effect and energy 
conservation (full case 1) are compared to: i) LPM effect without exact energy
conservation (case 2), ii) incoherent limit without energy conservation (case 3), iii)
incoherent limit with energy conservation (case 4), and iv) full case 1 with constraint 
that momentum transfer per scattering center is limited to $q_\perp < 2\mu$
(soft case 1). The bottom plot zooms into the small $L$-region.}
\label{fig1}
\end{figure}

We have calculated the average energy loss $\Delta E$ for a projectile quark of initial 
energy $E_q = 100$ GeV traversing an in-medium 
path length $L$. Results for the full MC algorithm, including LPM-effect and exact 
energy-momentum
conservation at each vertex are shown as '\underline{full case 1}' in Fig.~\ref{fig1}. 
All simulations are done for an elastic cross section 
 $\frac{d\sigma^{qQ_T\to qQ_T}}{d{q}_\perp^2} \sim \frac{1}{(\mu^2 + q_\perp^2)^2}$
with $\mu = 1\, {\rm GeV}^2$. We recall that the BDMPS multiple soft scattering 
approximation neglects the high-$q_\perp$ tail of this distribution~\cite{Baier:1996sk}. 
To mimic this approximation, we consider also a '\underline{soft case 1}' , 
in which the transverse momentum
range is $q_\perp \in [0,2\mu]$ rather than $q_\perp \in [0,\omega]$.

\begin{figure}[t]
\begin{center}
\includegraphics{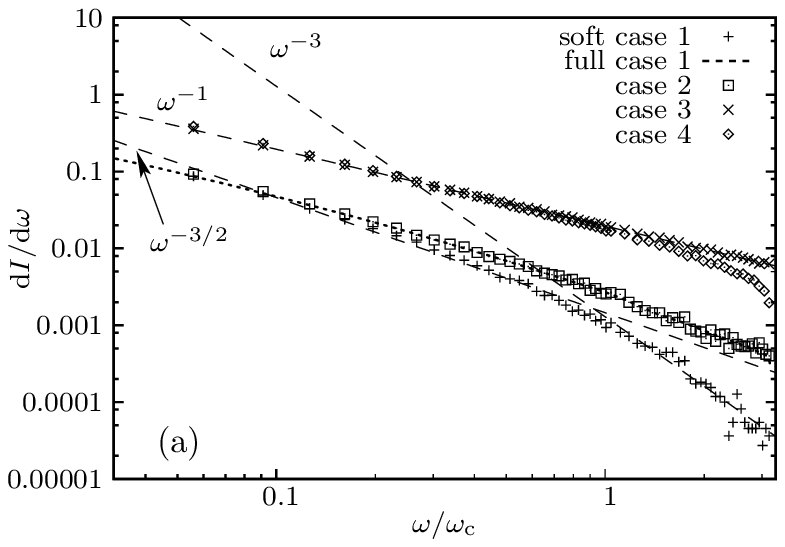}
\includegraphics{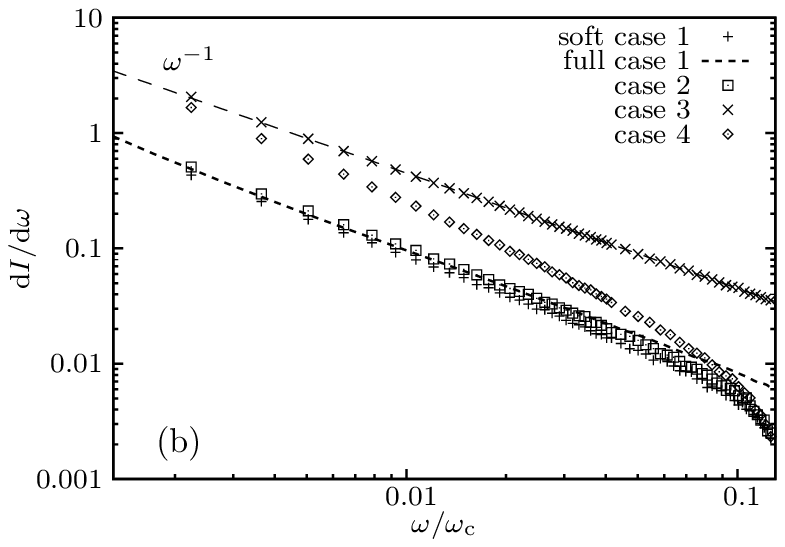}
\end{center}
\vspace{-0.5cm}
\caption{The medium-induced gluon energy distribution radiated from a quark of
energy $E_q = 100\, {\rm GeV}$ for the four cases of Fig.~\ref{fig1} and path length
(a) $L/L_c \approx 0.5$, (b) $L/L_c \approx 2.5$.
}
\label{fig2}
\end{figure}

For small $L$, results for the soft case 1 show an exact $\Delta E \propto L^2$-behavior 
(see Fig.~\ref{fig1}b), and the gluon spectrum shows the characteristic 
$\omega \frac{dI}{d\omega} \sim 1/\sqrt{\omega}$. [We find 
$dI/d\omega \propto 1/{\omega}^{1.45}$ for $\omega < 30\, {\rm GeV}$, data not shown.] 
This demonstrates that the MC algorithm reproduces the characteristic
BDMPS results if the soft BDMPS approximation is implemented. 
If one removes this soft approximation, $q_\perp \in [0,\omega]$,
one finds $\Delta E \propto L^{1.5}$, which lies in between the quadratic
dependence of coherent and the linear dependence of incoherent radiation. This is
so since large momentum transfers $\vert q_\perp \vert > 2\mu$ can shorten the
formation time  abruptly, thus introducing for small $L$ an incoherent component of gluon 
radiation. For the same reason, we find for the full case 1 and $\omega < 30\, {\rm GeV}$ 
a best fit to the spectrum $dI/d\omega \propto 1/{\omega}^{\beta}$, 
where $\beta = 1.38$ lies slightly further away from the BDMPS-limit
($\beta = 3/2$), and closer to the incoherent limit ($\beta = 1$), see Fig.~\ref{fig2}.

For sufficiently large $L> L_c =  \sqrt{\frac{4 \omega_{\rm max}}{ \hat{q}}}$,
a complete break-down of coherence is expected to occur, since
the kinematic constraint $\omega < \omega_{\rm max}$ implies a bound on the 
average formation time $\langle t_f \rangle < 2 \omega_{\rm max} / \langle k_T^2 \rangle$.
Within segments of length $L < L_c$, gluons can be emitted in a completely coherent 
process. But different segments of length $L_c$ add incoherently, and hence $\Delta E$ 
turns from a quadratic to a linear dependence around $L \sim L_c$, 
consistent with ~\cite{Baier:1996sk}. 
As $\Delta E$ grows with $L$, energy conservation
becomes more important 
and $\Delta E$ approaches $E_q$ for very large $L$, see Fig.~\ref{fig1}. 

In \underline{case 2}, we study the high-energy approximation that the projectile energy
is so large that its reduction due to gluon emission can be neglected. 
For small $L$, when energy conservation is unimportant, results on average energy loss
and the gluon distribution agree with case 1, see Fig.~\ref{fig1}b and ~\ref{fig2}a. 
For large $L$, however, since the projectile energy is kept fixed by assumption, all segments 
of length $L_c$ contribute the same, and $\Delta E \propto L$ as $L \gg L_c$
(see Fig.~\ref{fig1}a). Moreover, as for the incoherent limit, $dI/d\omega \sim 1/\omega$ 
for very large path length $L / L_c \gg 1$ (data not shown), but the gluon yield remains 
strongly suppressed in comparison to the incoherent limit, since scattering centers within
a distance $L_c$ act coherently.

We finally compare these results to the incoherent limits \underline{case 3} and 
\underline{case 4}, obtained by setting $t_f = 0$ in all intermediate steps of the MC algorithm. In 
case 3, the energy of the projectile is kept fixed as in case 2, i.e. exact energy 
conservation is abandoned, but one requires $\omega < \omega_{\rm max} = E_q$.
The average energy loss is then linear for arbitrary path length $L$ (see Fig.~\ref{fig1}).
For our model of the target, $dI/d\omega \sim 1/\omega$ (see Fig.~\ref{fig2}), and the slope 
of the linear dependence of $\Delta E \propto L$ depends logarithmically on the 
UV-cut-off $\omega_{\rm max}$. In case 4, we repeat the calculation with
exact energy conservation. For small $L$, this constraint is negligible, and results agree
with those of case 3. For large $L$,  however, energy loss is dominated by energy conservation 
and $\Delta E$ approaches $E_q$. Also, energy conservation implies that at large $L$, only soft
gluons can be emitted, since the projectile has lost already most of its energy. 
As a consequence, the spectrum at large $L$ tends to be much softer than 
$dI/d\omega \propto 1/\omega$. [The best fit in Fig.~\ref{fig2}b is 
$dI/d\omega \propto 1/{\omega}^{1.6}$, remarkably close to the LPM-spectrum.] 

%

To sum up: In vacuum, the dominant
quantum interference between subsequent parton branchings can be taken into account
probabilistically by angular ordering. Here, we have shown that the corresponding
dominant medium-induced quantum interference can be taken into account probabilistically 
by a formation time constraint. In particular, this reproduces by construction the analytically
known limiting cases of totally coherent and incoherent gluon production, and it
reproduces the characteristic $\Delta E \propto L^2$- and 
$\omega dI/d\omega \propto 1/\sqrt{\omega}$-dependencies of the non-abelian LPM effect, which were calculated in the soft scattering 
approximation~\cite{Baier:1996sk} (soft case 1), and it implements 
the known scale of breakdown of coherence effects~\cite{Baier:1996sk} at $L_c$.
By comparing simulations with medium-induced quantum interference to the incoherent 
limits (cases 3 and 4), our study also further quantifies the known importance of the LPM-effect
in the calculation of radiative parton energy loss.

Reproduction of known analytical limits is an important test for a MC algorithm of the 
LPM-effect. Finally, however, a major use of MC algorithms lies in exploring physics effects, 
which are difficult to treat in analytical formulations. The proposed MC algorithm
is suited to extend analytical treatments in several directions:
First, the analytical results of ~\cite{Baier:1996sk,Zakharov:1997uu,Wiedemann:2000za,Gyulassy:2000er,Arnold:2002ja,Wang:2001ifa} rely on 
approximations, in which energy is not conserved exactly,
and the energy of the incoming projectile remains unchanged in subsequent gluon emissions. 
In the present study, this corresponds to case 2, and it is seen to differ 
significantly from a formulation with exact energy-momentum conservation (case 1) outside
a region of relatively small path length and very high projectile energy.
Second, the analytical calculations of ~\cite{Baier:1996sk,Zakharov:1997uu,Wiedemann:2000za,Gyulassy:2000er,Arnold:2002ja,Wang:2001ifa} do not consider
the possibility that emitted gluons undergo further
medium-induced branchings. To compare to existing analytical 
results in Fig.~\ref{fig2}, we have adopted the same assumption here. But the MC
algorithm is easily extended to treat all incoherent components of a parton shower on equal
footing and to simulate the physically expected, further inelastic interactions of fully formed gluons. 
In addition, we have chosen here the ansatz  $\frac{d}{d\omega}\sigma^{qQ_T \to qQ_T g} \propto 1/\omega$, since it is inherent in the existing analytical treatments of the LPM effect,
and since it serves illustrative purposes (we wanted to have in the incoherent limiting case a $1/\omega$ power law, on top of which the characteristic interference-induced $\sqrt{\omega}$-modification is identified easily). The proposed MC algorithm, however, applies
equally well to arbitrary choices for $\sigma^{qQ_T\to qQ_T}$ and $\sigma^{qQ_T \to qQ_T g}$,
and allows us for instance to include a more realistic, steeper $\omega$-dependence of  
$\sigma^{qQ_T \to qQ_T g}$ close to the kinematic boundary $\omega \simeq E_q$. 
Moreover, realistic inelastic cross sections provide information about the recoil
of the target partons $Q_T$. The present proposal may open a new approach to follow
the dynamical evolution of these recoils. Another important challenge of future work is to
interface the present proposal with the virtuality evolution of a vacuum parton shower. 
We plan to pursue these questions in the context of the MC code JEWEL~\cite{Zapp:2008gi}.


\end{document}